\documentclass{ws-ijmpa}

\usepackage[english]{babel}

\usepackage[super,compress]{cite}

\usepackage{amsmath}
\usepackage{graphicx}
\usepackage{float}
\usepackage{url}

\usepackage[colorlinks=true, allcolors=blue]{hyperref}

\begin{document}
\renewcommand{\bibname}{References} 

\markboth{Abasov E., Boos E., Bunichev V., et.al.}{Angular correlations in associated top-quark and dark matter production at Large Hadron Collider}
%
\catchline{}{}{}{}{}
%
\title{Angular correlations in associated top-quark and dark matter production at Large Hadron Collider}
\author{Abasov E., Boos E., Bunichev V., Volkov P., Vorotnikov G., Dudko L., Zaborenko A., Iudin E., Markina A., Perfilov M.}

\address{Skobeltsyn Institute of Nuclear Physics, Lomonosov Moscow State University, Leninskie gory, GSP-1\\
Moscow 119991, Russian Federation\\
emil@abasov.ru}
\maketitle

\begin{history}
\received{}
\revised{}
\end{history}

\begin{abstract}
In the paper, we consider the processes of single top-quark production in association with dark matter particles through the t-channel in simplified models with scalar and pseudoscalar mediators. Within the framework of these models, we analyze the angular correlations that arise in the top quark production process in association with dark matter, comparing them with the production of a top quark in the Standard Model. We consider angular correlations of kinematic variables, which define admissible regions in momentum space and can be used experimentally to distinguish between the fraction of the total momentum carried away by the scalar or pseudoscalar mediator. We demonstrate that scalar mediator significantly changes the rest frame of the well-known angular correlation observable and changes the distributions. In this case, this variable provides clear separation between the SM and DM contribution and can be used to increase the sensitivity of future DM analyses.
\end{abstract}

\ccode{PACS numbers: 14.65.Ha, 95.35.+d, 07.05.Mh}
\newpage
\tableofcontents
\section{Introduction}
Convincing astrophysical and cosmological observations indicate
the presence of new, weakly interacting, neutral substance called Dark Matter (DM). Under the assumption that dark 
matter has the nature of particles, various extensions of the Standard Model (SM) are created to explain the origin of DM and its possible detection in an experiment.

Direct searches for dark matter can be performed in the processes of interaction
between cosmic dark matter particles and Earth's detectors~\cite{LUX:2012kmp}. At the same time, the Large Hadron Collider (LHC) presents a unique opportunity to observe and study the production of dark matter particles under laboratory conditions. This possibility is based on calculations of the cross sections for the production of hypothetical dark matter particles, which, as it turns out, may be on the order of the cross sections for processes caused by weak interactions at LHC energies~\cite{Buckley:2011kk}.
Since dark matter particles are not detectable in the LHC detectors, the only way to detect them is if they were produced along with the visible particles described by the SM. In that case, the dark matter particle in its final state carries away energy, which is defined in the detector as missing energy.

The possible interaction of dark matter particles with SM particles can be realized by means of an intermediate particle, the so-called mediator. Depending on the considered mechanism of interaction between dark matter particles and SM particles, the mediator is endowed with different properties. 
In the case of very heavy mediators, a possible type of interaction could be a contact interaction involving a quark-antiquark pair or two gluons and two dark matter particles. In this case, the magnitude of the signal and its distribution on missing energy are determined by the nature and mass of dark matter particles, as well as the Lorentz structure and interaction strength. To model such scenarios and calculate the magnitudes of possible signals, one uses contact interaction operators of effective field theory (EFT)~\cite{Goodman:2010yf}, also called DMEFT. 

In the case of a presumably non-heavy mediator, models are needed that explicitly include such mediators in consideration. Such models, called simplified models~\cite{Alwall:2008ag,Abdallah:2015ter,LHCNewPhysicsWorkingGroup:2011mji,Abercrombie:2015wmb}, are also intensively used in experimental searches for DM particles at the LHC. Although the total number of these models is quite large, models in which dark matter particles interact with SM particles through new scalar, pseudoscalar or vector mediators are of particular interest due to their simplicity.

Assuming minimal flavour violation (MFV)~\cite{Chivukula:1987py,Hall:1990ac,Buras:2000dm,DAmbrosio:2002vsn}, third-generation quarks may play a significant role~\cite{Lin:2013sca} in models with scalar and pseudoscalar mediators.
Therefore, it is of particular interest to consider collider processes of DM production in association with the top quark which is the heaviest one. 

This idea has motivated experimental searches for events in which DM particles are produced in association with a pair of top quarks ($t \bar t + DM$)~\cite{ATLAS:2014dbf,CMS:2016mxc} or with a single top quark ($t/\bar t +DM$)~\cite{Pinna_2017}. As a result of a search in LHC experiments for such processes no significant excess of the signal was found above the predicted background of the Standard Model~\cite{CMS:2019zzl,ATLAS:2022znu,ATLAS:2020yzc}.

The presence of non-SM particles can change the known distributions of kinematic variables in the SM. In single top production, the top quark is highly polarized due to (V-A) electroweak vertices in production and decay of the top quark. The $P_z$ component of the polarization vector is proportional to the cosine of the angle between the spectator quark and the charged lepton in the top-quark decay chain, in the rest frame of the top-quark~\cite{PhysRevD.55.7249}. This angular correlation is a well-known property and is very sensitive to deviations in Wtb interactions. 
The production of scalar dark matter mediator from the top-quark-line
does not change the polarization of the top-quark, but the direction of
the d-quark momentum is not now the axis along which the top quark is
about 100\% polarized, and as a result, changes the observable cosine.

In this work, we analyze angular correlations in the single top quark production  in association with a scalar and pseudoscalar DM mediators. It would seem obvious that the kinematic distributions associated with angular correlations should not change, at least in the case of a scalar mediator. However, the presented analysis shows that this is not the case. An example of a kinematic variable associated with angular correlations, which is affected by a mediator, including a scalar one, is given.
A number of proposed kinematic variables are used to train a neural network that separates the DM mediator and SM neutrino contributions to the total missing energy determined at the detector.
It is shown that multivariate analysis techniques such as deep neural networks can help improve the sensitivity of the analysis to the presence of DM mediators. In the framework of the considered simplified models with various mediators, in this article we focused on the spin properties of the t-channel single top quark production in association with the DM mediators.

The paper is organized as follows. Section~\ref{sec:models} describes two classes of phenomenological models that predict the processes of top-quark production in association with DM particles and gives an overview of the latest results of the search for DM particles in association with top-quarks. 
The section~\ref{sec:spincor} describes and summarizes angular correlations in the processes of associative production of dark matter with top quarks at the LHC. In Section~\ref{sec:newdev} the analytical expressions for mediator momentum components are provided and additional methods of improving the manifestation of angular correlations are discussed. The section~\ref{sec:results} gives the final results. The conclusion describes possible applications of the proposed method for increasing the sensitivity for DM searches at the LHC. 

The computational package CompHEP~\cite{CompHEP:2004qpa,Pukhov:1999gg} was used for the simulations and numerical calculations. All calculations are given at a proton-proton collision energy of 14 TeV for the Large Hadron Collider (LHC).

\section{Simplified models with scalar and pseudoscalar mediators predicting the processes of top-quark production in association with DM particles}
\label{sec:models}
\begin{figure}[ht]
    \centering
    \includegraphics[width=0.35\linewidth]{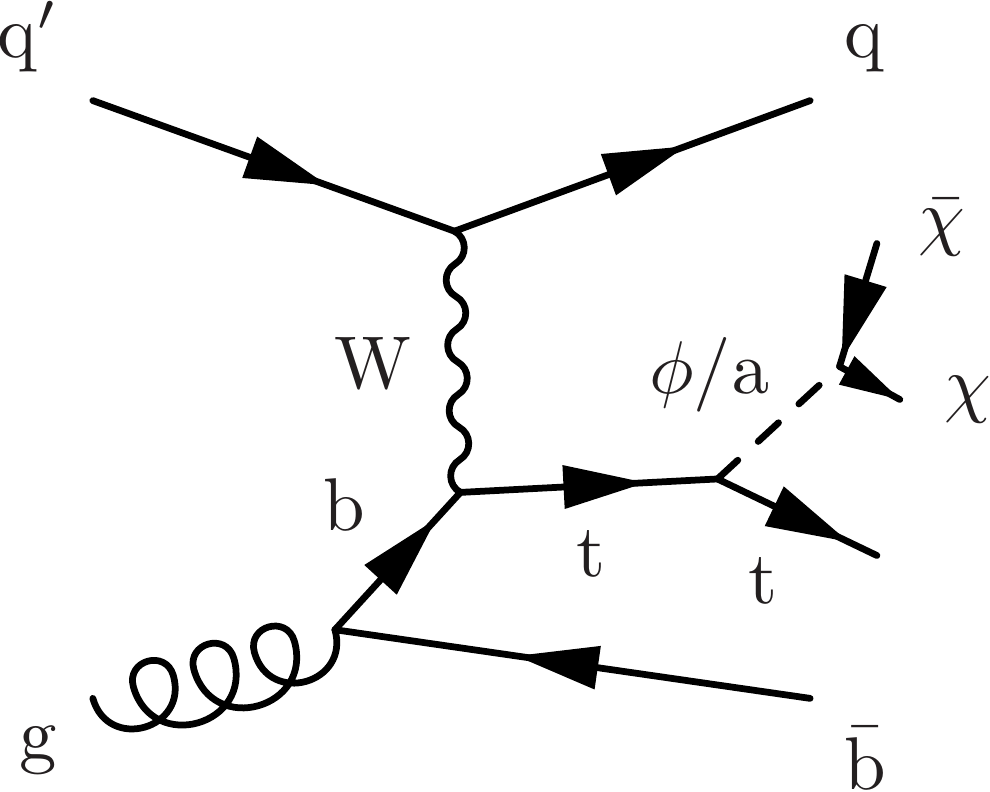}
    \caption{Representative Feynman diagram of the $q'g \rightarrow t\Bar{b}q$ process with the presence of scalar/pseudoscalar mediator.}
    \label{fig:diagram}
\end{figure}

For simplified DM models it is assumed that DM particles interact with SM particles by exchanging one or more particles, called ``mediators'', which have weak coupling with SM particles. Through the interactions of the mediators, a so-called ``portal'' for the interaction between the SM and DM particles is provided, leading to effective interactions beyond the SM. This leads to the possibility of the production of SM particles together with DM particles and, accordingly, to the observation of the characteristic ``missing energy'' in such processes, since DM particles are not directly detected. The joint production of SM and DM particles, with the subsequent detection of the ``missing energy'' is the only way to detect weakly interacting dark matter in experiments at high-energy colliders and experiments with disembodied beams, which is the reason for the relevance of the topic of this section.
In spite of the fact that there are many different variants of simplified models, at present those in which dark matter particles interact with SM particles by means of new scalar, pseudoscalar, or vector mediators look theoretically attractive. 
Within the framework of the considered simplified models with different mediators there are processes of DM production in association with a single top-quark, a pair of top-antitop-quarks, with three and four top-quarks. Further in this paper all methods are considered on the example of the process with the t-channel production of a single top-quark.
In simplified models, dark matter particles are assumed to be Dirac fermions with the interaction between SM and dark matter sectors occurring via either a massive electrically neutral scalar or a pseudoscalar particle, often denoted by a single symbol $\phi$ and A correspondingly. 
Interaction Lagrangians between scalar/pseudoscalar particles and SM fermions and dark matter are written as:
\begin{equation}
L_{\Phi}=g_\chi \Phi \bar{\chi} \chi+\frac{g_v \Phi}{\sqrt{2}} \sum_f\left(y_f \bar{f} f\right) 
\end{equation}
\begin{equation}
L_A=i g_\chi A \bar{\chi} \gamma^5 \chi+i \frac{g_v A}{\sqrt{2}} \sum_f\left(y_f \bar{f} \gamma^5 f\right)
\end{equation}
Typical Feynman diagram of the studied process is presented in the Figure~\ref{fig:diagram}.
\section{Angular correlations in associated production of top-quark and dark matter at LHC}
\label{sec:spincor}

\begin{figure}
    \centering
    \includegraphics[width=.9\linewidth]{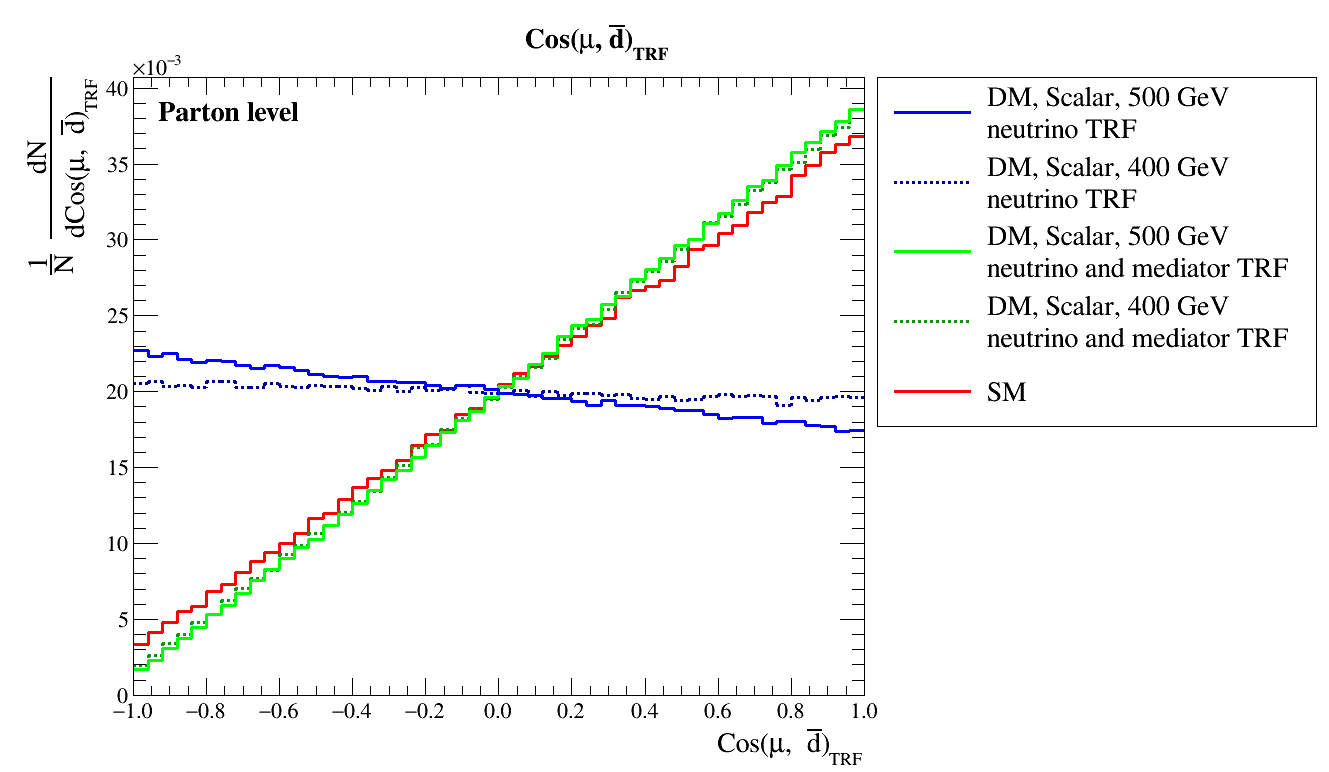}
    \caption{Distribution of cosine of angle between lepton and down-type quark in top-quark rest frame in processes $pp\rightarrow t(\rightarrow \nu_l\Bar{l}b)q\ (SM)$ and $pp\rightarrow t(\rightarrow \nu_l\Bar{l}b)q\Bar{\chi}\chi\ $ (``neutrino'' TRF) with presence of a scalar mediator in a model with $m_\chi = 1\ GeV$, 2 samples with $m_\phi =400\ GeV$ and $500\ GeV,\ g_\chi=g_\nu=1$. For comparison the same distribution in a top-quark emitting the DM mediator is also provided (``neutrino and mediator'' TRF).}
    \label{fig:spincorrs_parton_scal}
\end{figure}

\begin{figure}
    \centering
    \includegraphics[width=.9\linewidth]{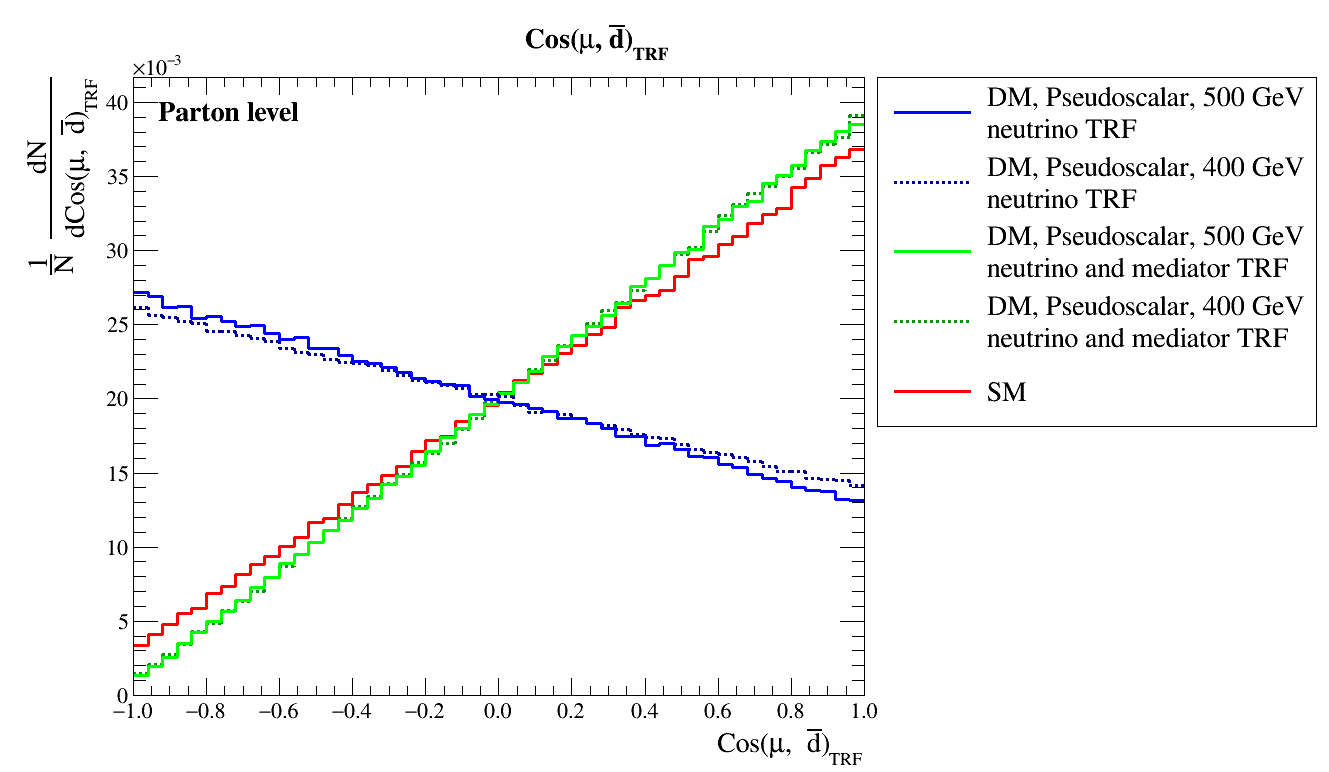}
    \caption{Distribution of cosine of angle between lepton and down-type quark in top-quark rest frame in processes $pp\rightarrow t(\rightarrow \nu_l\Bar{l}b)q\ (SM)$ and $pp\rightarrow t(\rightarrow \nu_l\Bar{l}b)q\Bar{\chi}\chi\ $(neutrino TRF) with presence of a pseudoscalar mediator in a model with $m_\chi = 1\ GeV$, 2 samples with $m_\phi =400\ GeV$ and $500\ GeV,\ g_\chi=g_\nu=1$. For comparison the same distribution in a top-quark emitting the DM mediator is also provided (neutrino and mediator TRF)).}
    \label{fig:spincorrs_parton_pseudoscal}
\end{figure}

This article focuses on single-top quark production with subsequent leptonic decay, which is well studied in the SM. One of the significant properties of such process in SM is correlation of the spin of top-quark and its decay products with the direction of the down-type quark. From that one can get the relation for the cosine of angle between down-type quark and the lepton in a specific case of $ug \rightarrow t\Bar{b}d$ and a subsequent $t \rightarrow \Bar{l}\nu b$ decay~\cite{PhysRevD.55.7249}. Since the mediator is a scalar particle, changes to this rule were not expected. For the top-quark rest frame (TRF), firstly, 4-momentum of the top-quark is calculated: as a sum of 4-momenta of neutrino, charged lepton and the b-quark for the ``neutrino'' TRF case and with the addition of mediator in the ``neutrino and mediator'' TRF case. Secondly, charged lepton and  down-type quark are boosted to the TRF using the previously discussed top-quark momentum. Then, the cosine of the angle between the charged lepton and down-type quark is calculated as the normalized dot product of corresponding momenta.
Thus, the distribution changes due to the inclusion of the heavy mediator in the TRF calculations in the first step.
\begin{equation}
 |M^2|\propto m_t^2 E_{\bar{d}} E_{\bar{\ell}}\left(1+\cos \theta_{\bar{d} \bar{\ell}}\right)
\end{equation}
 To check this property events with DM mediator were generated in CompHEP using a scalar mediator model. However, as can be seen on Fig.~\ref{fig:spincorrs_parton_scal}, addition of the DM mediator breaks the angular correlations. This is caused due to the cosine being constructed in the top-quark rest frame (TRF), which is changed with the additional emission of a heavy scalar boson. Since all particles are available at a parton level generation, we have added the mediator to the reconstruction of the missing transverse energy, which restored the expected distribution.
The same algorithm was used for events with pseudoscalar mediator model. As can be seen in Figure~\ref{fig:spincorrs_parton_pseudoscal}, it displays the same general behaviour to the sample with the scalar mediator.

\begin{figure}
    \centering
    \includegraphics[width=.9\linewidth]{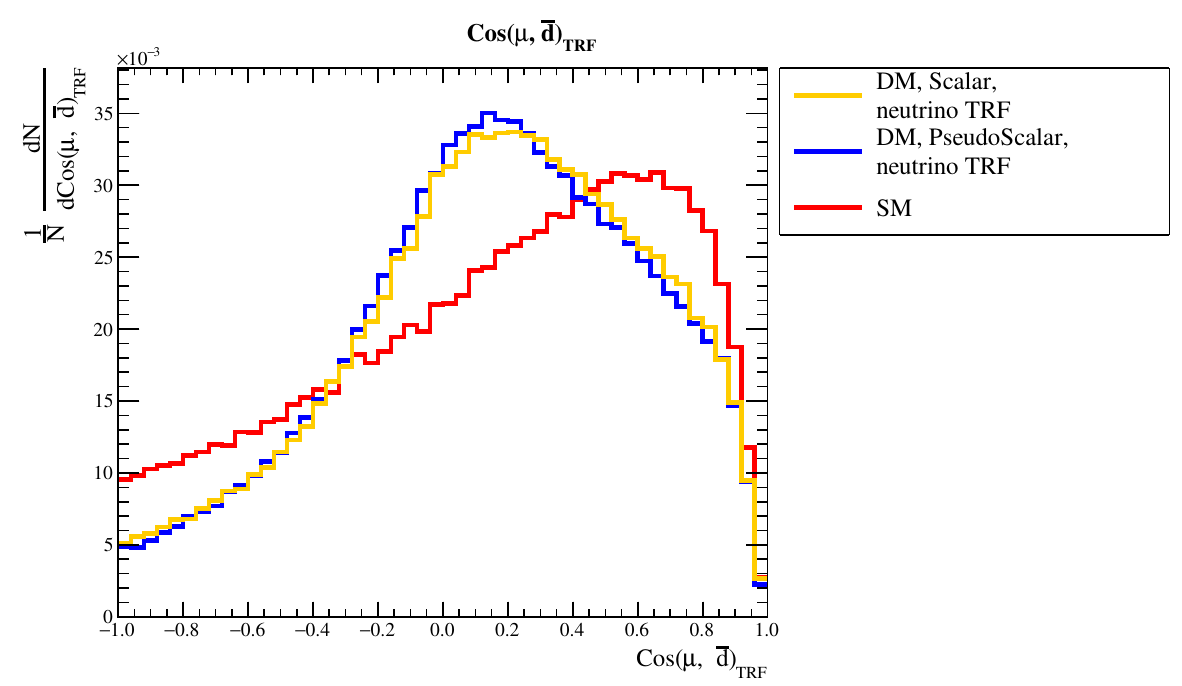}
    \caption{Distribution of cosine of angle between lepton and down-type quark in top-quark rest frame with simulation of detector response in Delphes for the processes $pp\rightarrow t(\rightarrow \nu_l\Bar{l}b)q\ (SM)$ and $pp\rightarrow t(\rightarrow \nu_l\Bar{l}b)q\Bar{\chi}\chi\ (DM)$ with presence of a scalar/pseudoscalar mediator in a model with $m_\chi = 1\ GeV,\ m_\phi =400\ GeV,\ g_\chi=g_\nu=1$. The rest frame is reconstructed as for usual SM analysis with assumption that all missing transverse energy comes from W boson in top quark decay.}
    \label{fig:spincorrs_delphes}
\end{figure}

Next, all events were hadronized in Pythia8~\cite{Bierlich:2022pfr} and run through a detector response simulation with Delphes~\cite{deFavereau:2013fsa}. The selection criteria for Delphes are: 1 lepton with $P_t > 26$ GeV, $\eta < 2.4$, 2 or 3 jets with $P_t > 30$ GeV, exactly one of which is b-tagged, $MET > 20$ GeV. As can be seen in Fig.~\ref{fig:spincorrs_delphes}, distributions still clearly differ, which allows an additional observable for the separation of DM events in single top-quark production based on angular correlations.

\section{Further developments, allowing better separation of processes with DM particles}
\label{sec:newdev}

\subsection{Analysis of the analytical properties of the process}
\label{subsec:analyt}
Patterns after Delphes simulation of detector smearing discussed in the previous sections were studied based on the assumption that all missing transverse momentum comes from neutrino, as it is used for SM measurements. 
Since the mediator has a significant influence on the top quark rest frame, it can be assumed that taking into account the mediator in reconstruction will greatly improve the separation between dark matter and standard model contributions. However, the results of detector simulations show that the mediator and neutrinos are only accounted for in the transverse missing energy, which is not sufficient to separate the contribution of the mediator. One possible way to achieve this is by using the kinematic properties of a given process.
For the single top-quark production these relations are the following.

Let the $P_\nu = (|\vec{p_{\nu}}|, p_{\nu_x}, p_{\nu_y}, p_{\nu_z}) ,P_\phi = (E_{\phi}, p_{\phi_x}, p_{\phi_y}, p_{\phi_z})$ be 4-momenta of neutrino and mediator.
From energy and momentum conservation laws one can get:
\begin{align}  
    (P_{l} + P_{\nu})^2 &= M_{w}^{2} \label{eq:1}\\
    (P_{l} + P_{\nu} + P_b)^2 &= M_{t}^{2}\label{eq:2} \\
    p_{\nu_x} + p_{\phi_x} &= MET_x \label{eq:3} \\
    p_{\nu_y} + p_{\phi_y} &= MET_y \label{eq:4} \\
    P_{\phi}^{2} &= M_{\phi}^{2}\label{eq:5}
\end{align}

\begin{figure}[H]
	\centering
    \includegraphics[width = .6\linewidth]{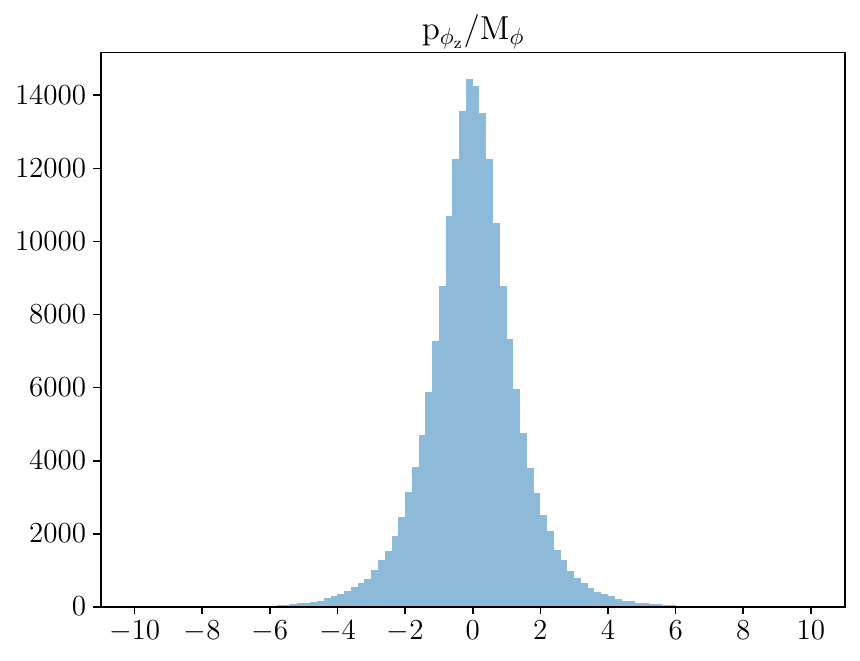}
	\caption{Distribution of $\frac{p_{\phi_z}}{M_{\phi}}$. On $y$ axis -- event count, on $x$ -- value of a variable.}
	\label{fig:pz}
\end{figure}

In this system equation~\eqref{eq:1} describes the $W \rightarrow l \nu$ decay, \eqref{eq:2} - $t \rightarrow W b$ decay. Equations~\eqref{eq:3} and~\eqref{eq:4} arise from the registered missing transverse energy (MET) components in the detector. Finally, \eqref{eq:5} is a known constraint for on-shell mediator 4-momentum based on its mass.
Rewriting \eqref{eq:5} with \eqref{eq:3},\eqref{eq:4} and using 4-momentum for $P_\phi$, one gets
\begin{equation}
    E_{\phi}^{2} - (MET_x - p_{\nu_x})^2 - (MET_y - p_{\nu_y})^2 - p_{\phi_z}^{2} = M_{\phi}^{2}
\end{equation}
Taking into account that $\frac{p_{\phi_z}}{M_{\phi}}$ is distributed close to 0 (Fig.~\ref{fig:pz}), one can omit the last term in the equation.

\begin{equation}
\left(\frac{E_{\phi}}{M_{\phi}}\right)^{2} - \left(\frac{MET_x - p_{\nu_x}}{M_{\phi}}\right)^2 - \left(\frac{MET_y - p_{\nu_y}}{M_{\phi}}\right)^2 - 1 = 0 
\end{equation}
Expanding the equations \eqref{eq:1} and \eqref{eq:2}:
\begin{equation}
\begin{aligned} 
& \left(E_l+\sqrt{p_{\nu_x}^2+p_{\nu_y}^2+p_{\nu_z}^2}\right)^2-\left((p_{\nu_ x}+p_{l_x})^2+(p_{\nu_y}+p_{l_y})^2+(p_{\nu_z}+p_{l_z})^2\right)=M_{w}^2,\\
&  \left(E_l+E_b+\sqrt{p_{\nu_x}^2+p_{\nu_y}^2+p_{\nu_z}^2}\right)^2-\left((p_{\nu_x}+p_{l_x} + p_{b_x})^2 +(p_{\nu_y}+p_{l_y}+ p_{b_y}\right)^2+ \\
&+(p_{\nu_z}+p_{l_z}+ p_{b_z})^2)=M_{t}^2 
&
\end{aligned}  
\end{equation}
Expanding the squares, omitting $M_l^2$ and introducing a new variable
$$A = -M_{t}^2 + 2E_{l}E_{b} + M_{b}^2 - 2(\vec{p_l},\vec{p_b}), $$
one gets 
\begin{equation}
  \begin{aligned}
   M_{w}^{2} + A + 2E_{b}|\vec{p_\nu}| - 2(\vec{p_\nu},\vec{p_b}) = 0  \\
   M_{w}^{2} - 2E_{l}|\vec{p_\nu}| + 2(\vec{p_\nu},\vec{p_l}) = 0 
  \end{aligned}
\end{equation}

Knowing that $\frac{E_b}{E_l}\left( M_{w}^{2} + 2 (\vec{p_\nu},\vec{p_l})\right) = 2 E_b |\vec{p_\nu}|$, the system can be rewritten as:
\begin{equation}
    \left\{\begin{array}{l}
M_w^2\left(1+\frac{E_b}{E_l}\right)+A+2\left(\vec{p_\nu}, \frac{E_b}{E_l} \vec{p_l}-\vec{p}_b\right)=0 \\ 
M_w^2-2 E_{l}\left|\vec{p_\nu}\right|+2\left(\vec{p_\nu}, \vec{p_l}\right)=0 
\end{array}\right. \\ 
\label{eq:10}
\end{equation}
Introduce $\tilde{A} = M_w^2\left(1+\frac{E_b}{E_l}\right)+A$, $\vec{p}_{1b} = 2\frac{E_b}{E_l} \vec{p_l}-2\vec{p_b}$. Then the first equation in~\eqref{eq:10} can be rewritten as 
\begin{equation}
\tilde{A} +p_{\nu_x} p_{1b_x}+p_{\nu_y} p_{1b_y}+p_{\nu_z} p_{1b_z}=0 
\end{equation}

and the solution for $z$-component of neutrino momentum can be achieved: 
\begin{equation}
p_{\nu_z}=-\frac{\tilde{A}+p_{\nu_x} p_{1b_x}+p_{\nu_y} p_{1b_y}}{p_{1b_{z}}}
\end{equation}
Let's consider the second equation in~\eqref{eq:10}:
\begin{equation}
M_{w}^2-2E_l\sqrt{p_{\nu_x}^2+p_{\nu_y}^2+p_{\nu_z}^2}+2p_{\nu_x} p_{l_x}+2 p_{\nu_y} p_{l_y}+2p_{\nu_z} p_{l_z}=0 
\end{equation}
Multiplying it by $p_{1b_{z}}$ and grouping the terms, one gets:
\begin{equation}
\begin{array}{l}
M_w^2 p_{1b_z}-2p_{l_z}\tilde{A}+p_{\nu_x}\left(2 p_{l_x} p_{1b_z}-2 p_{l_z} p_{1b_x}\right)+ \\
 +p_{\nu_y}\left(2p_{l_y} p_{1b_z}-2p_{l_z}p_{1b_y}\right)=2E_l p_{1b_z} \sqrt{p_{\nu_x}^2+p_{\nu_y}^2+p_{\nu_z}^2} 
 \end{array} 
\end{equation}
Introducing new variables $$C = \frac{M_w^2 p_{1b_z} - 2p_{l_z}\tilde{A}}{2E_l p_{1b_z}}, C_x = \frac{p_{l_x} p_{1b_z} - p_{l_z}p_{1b_x}}{E_l p_{1b_z}}, C_y = \frac{p_{l_y} p_{1b_z} - p_{l_z}p_{1b_y}}{E_l p_{1b_z}},$$ one gets
\begin{equation}
C + C_x p_{\nu_x} + C_y p_{\nu_y} = \sqrt{p_{\nu_x}^2+p_{\nu_y}^2+p_{\nu_z}^2}
\end{equation}
\begin{equation}
    \begin{array}{l}
    p_{\nu_x}^2+p_{\nu_y}^2+p_{\nu_z}^2 = p_{\nu_x}^2+p_{\nu_y}^2+ \left(\frac{\tilde{A}+p_{\nu_x} p_{1b_x}+p_{\nu_y} p_{1b_y}}{p_{1b_{z}}}\right)^2 = \\
    = p_{\nu_x}^2 \left(1 + \left(\frac{p_{1b_x}}{p_{1b_{z}}}\right)^2\right) + p_{\nu_y}^2 \left(1 + \left(\frac{p_{1b_y}}{p_{1b_{z}}}\right)^2\right) + \left(\frac{\tilde{A}}{p_{1b_{z}}}\right)^2 + \\
    2 \frac{\tilde{A} p_{1b_{x}}}{p_{1b_{z}}^2 } p_{\nu_x} + 2 \frac{\tilde{A} p_{1b_{y}}}{p_{1b_{z}}^2} p_{\nu_y} + 2 \frac{p_{1b_{x}} p_{1b_{y}}}{p_{1b_{z}}^2} p_{\nu_x} p_{\nu_y} =  C^2 + \\
     + (C_x p_{\nu_x})^2 + (C_y p_{\nu_y})^2 + 2 C C_x p_{\nu_x} + 2 C C_y p_{\nu_y}  + 2 C_x C_y p_{\nu_x} p_{\nu_y} 
     \end{array} 
\end{equation}
Some additional variables for simplicity:
 $$\begin{array}{l}
\tilde{C} = C^2 - \left(\frac{\tilde{A}}{p_{1b_{z}}}\right)^2, \tilde{C_{x^2}} = C_x^2 - \left(1 + \left(\frac{p_{1b_x}}{p_{1b_{z}}}\right)^2\right), \tilde{C_{y^2}} = C_y^2 - \left(1 + \left(\frac{p_{1b_y}}{p_{1b_{z}}}\right)^2\right), \\
\tilde{C_{xy}} = C_x C_y - \frac{p_{1b_{x}} p_{1b_{y}}}{p_{1b_{z}}^2}, \tilde{C_{x}} = C C_x - \frac{\tilde{A} p_{1b_x}}{p_{1b_z}^2}, 
\tilde{C_{y}} = C C_y - \frac{\tilde{A} p_{1b_y}}{p_{1b_z}^2}
\end{array} $$
Finally, one gets the equation:
\begin{equation}
\tilde{C} + \tilde{C_{x^2}} p_{\nu_x}^2 + \tilde{C_{y^2}} p_{\nu_y}^2 + 
2 \tilde{C_{xy}} p_{\nu_x} p_{\nu_y} + 2 \tilde{C_{x}} p_{\nu_x} +  2 \tilde{C_{y}} p_{\nu_y} = 0
\end{equation}
Which can be solved for $p_{\nu_x}$: 
\begin{equation}p_{\nu_x} = \pm \sqrt{d p_{\nu_y}^2 + e p_{\nu_y} + f} - \frac{\tilde{C_x}}{\tilde{C_{x^2}}} - \frac{\tilde{C_{xy}}}{\tilde{C_{x^2}}} p_{\nu_y} 
\end{equation}
where $$d = \frac{\tilde{C_{xy}^{2}} - \tilde{C_{x^2}} \tilde{C_{y^2}}}{\tilde{C_{x^2}}^2}, e = \frac{2 \tilde{C_{xy}}\tilde{C_{x}} - 2 \tilde{C_{x^2}} \tilde{C_{y}}}{\tilde{C_{x^2}}^2}, f = \frac{\tilde{C_{x}^2} - \tilde{C} \tilde{C_{x^2}} }{\tilde{C_{x^2}}^2}$$
This is a surface $p_{\nu_x}(p_{\nu_y})$.

No simple relation between mediator and neutrino momenta could be obtained from the above relations, like it was achieved for two neutrinos in pair top-quark production~\cite{Betchart:2013nba}. So, the equations provide little insight into the improvement in the separation of neutrino and mediator momenta. Additional approximations are needed in order for the more clear dependencies to appear.

\subsection{Neutrino momentum reconstruction with neural networks}
\label{subsec:NN}
Another opportunity to utilize angular correlations lies in learning the reconstruction of neutrino and DM mediator momenta from the MC simulation directly. This task can be solved with the application of various machine learning methods.
For this work, we have trained a Multi-Layer Perceptron in a multi-output regression setting for two separate test cases: 
\begin{itemize}
    \item A network trained only on SM data, referred to as the SM network, with 3 output variables: $p_x, p_y, p_z$ components of neutrino momentum
    \item A complete network, trained on the mix of SM and DM events, referred to as the SM+DM network, with 5 output variables: $p_x, p_y, p_z$ components of momenta for neutrino, $p_z$ component for the DM mediator as well as the mass of the mediator. $p_x$ and $p_y$ components of momenta of the mediator are calculated using relations \eqref{eq:3} and \eqref{eq:4} to simplify the neural network.
\end{itemize}

The reconstructed variables are used to calculate the cosine of the angle between the lepton and the down-type quark in the top-quark rest frame. 
Apart from the difference in output dimensions, the 2 networks are identical. For the input variables 3-momenta of the lepton, b-quark and down-type quarks are used. In addition to them, some high-level variables are constructed following the approach described in Ref.~\citen{Boos:2008sdz,Dudko:2020qas}.

The Standard Scaler is used for all input and output variables; in order to maintain a fair comparison, all output variables are inversely scaled after the inference, and are compared with the original. We use Mean Absolute Percentage Error as the tuning criterion. The resulting network is trained until the validation loss stops decreasing.
\subsubsection{SM network}
\begin{figure}[h]

    \includegraphics[width=0.32\textwidth]{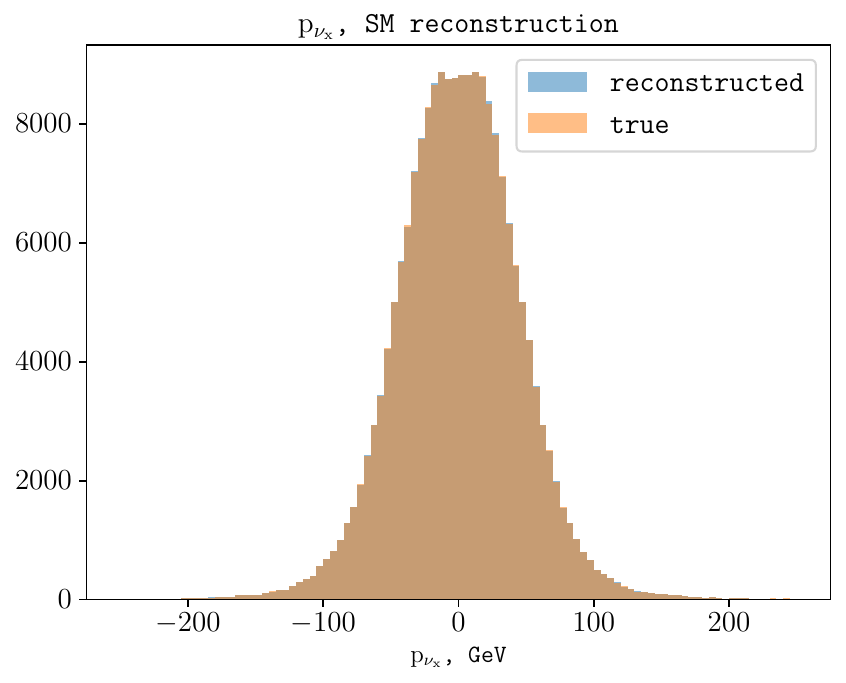} 
    \includegraphics[width=0.32\textwidth]{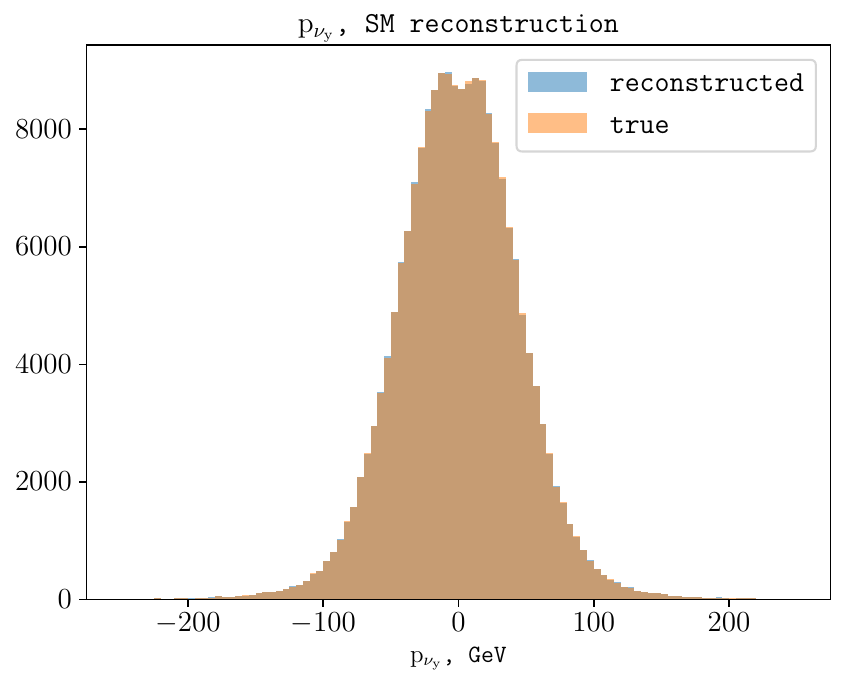}
    \includegraphics[width=0.32\textwidth]{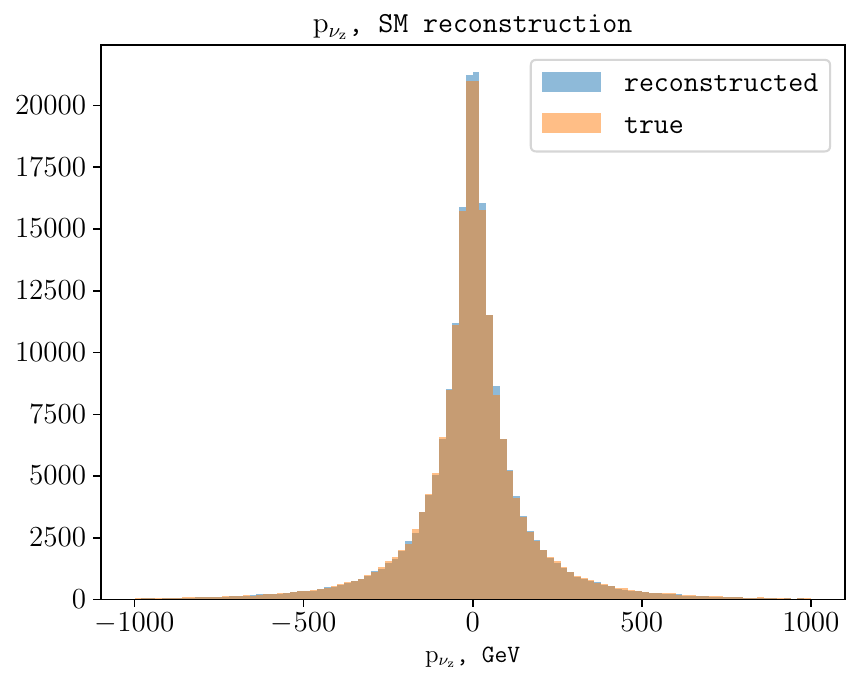}

\caption{Event distributions for the outputs of the network trained on SM sample only. Predicted values shown in blue, target values - in orange.}
\label{fig:SM_net}
\end{figure}

\begin{figure}[h]
    \centering
    \includegraphics[width=.5\textwidth]{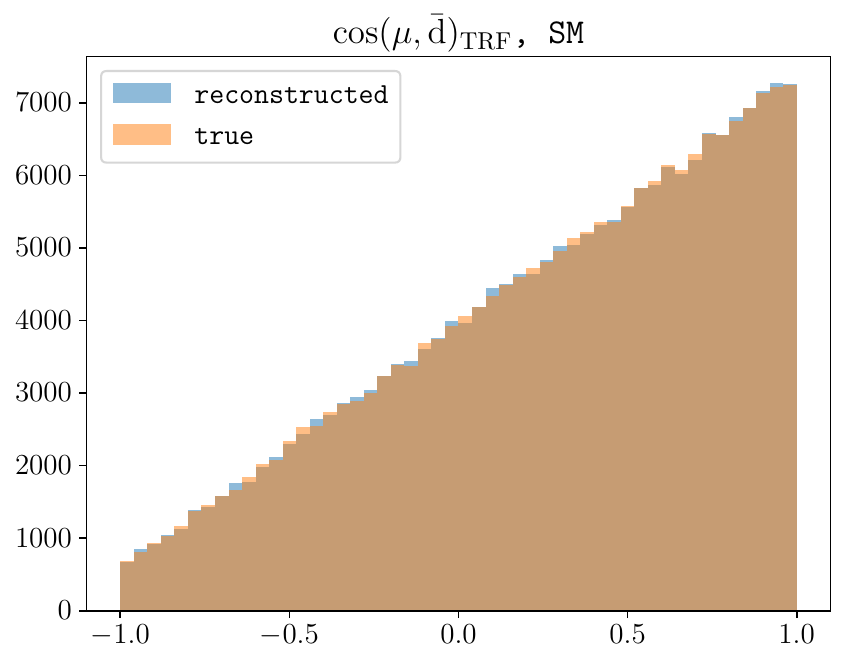} 
\caption{Event distributions for the calculated $cos\theta_{\bar{d} \bar{\ell}}$ using reconstructed neutrino momentum from the SM NN.}
\label{fig:SM_net_cos}
\end{figure}

For the SM network only the SM sample from CompHEP is used. This is intended as a proof-of-concept case for our method of momentum reconstruction.
As can be seen from Fig.~\ref{fig:SM_net} and~\ref{fig:SM_net_cos}, reconstructed distributions of both the components of the neutrino momentum and $cos\theta_{\bar{d} \bar{\ell}}$ perfectly align with the desired values obtained from MC.

\subsubsection{SM+DM network}
\begin{figure}[h]

    \includegraphics[width=0.32\textwidth]{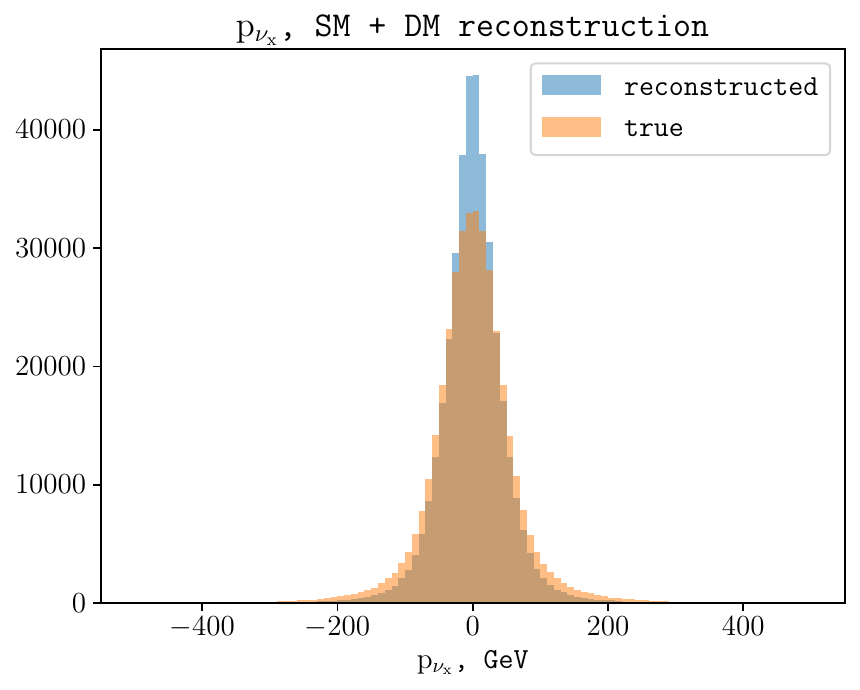} 
    \includegraphics[width=0.32\textwidth]{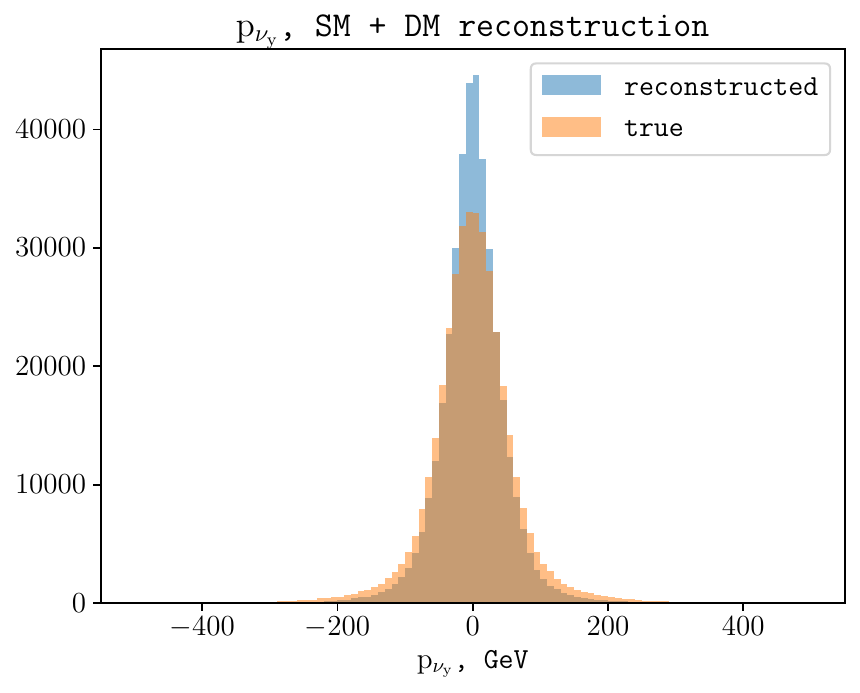}
    \includegraphics[width=0.32\textwidth]{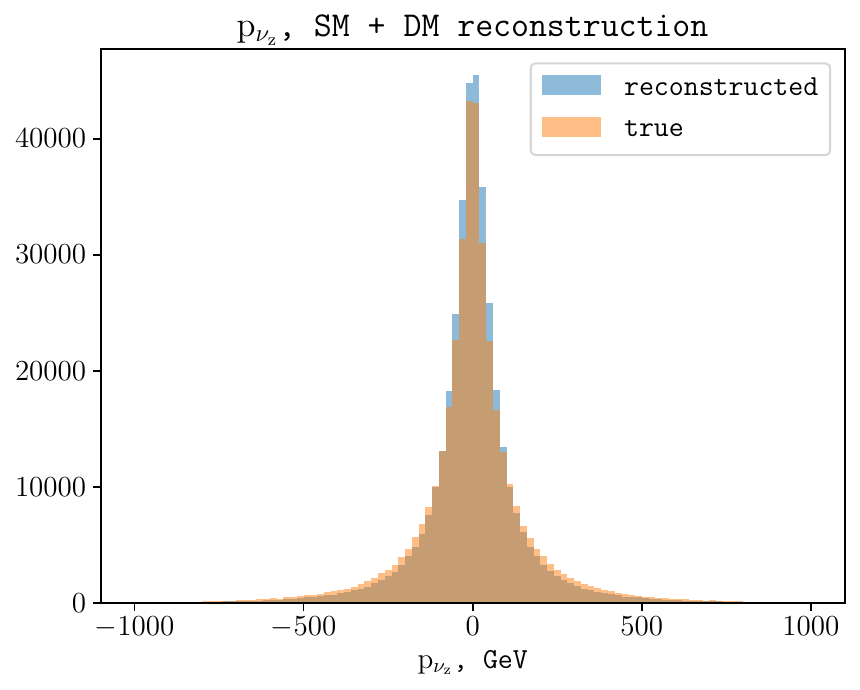} 
    \includegraphics[width=0.32\textwidth]{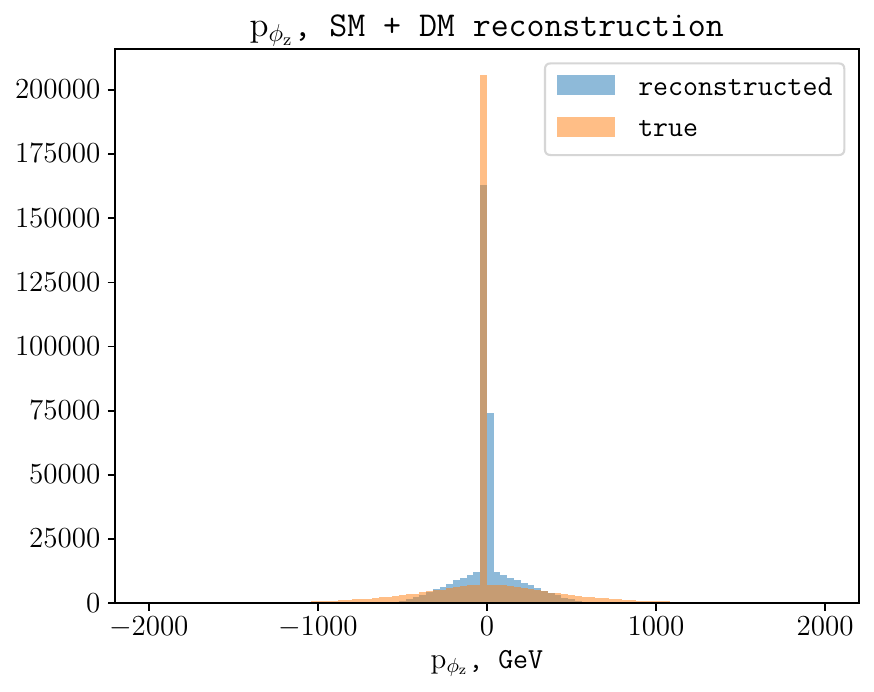} 
    \includegraphics[width=0.32\textwidth]{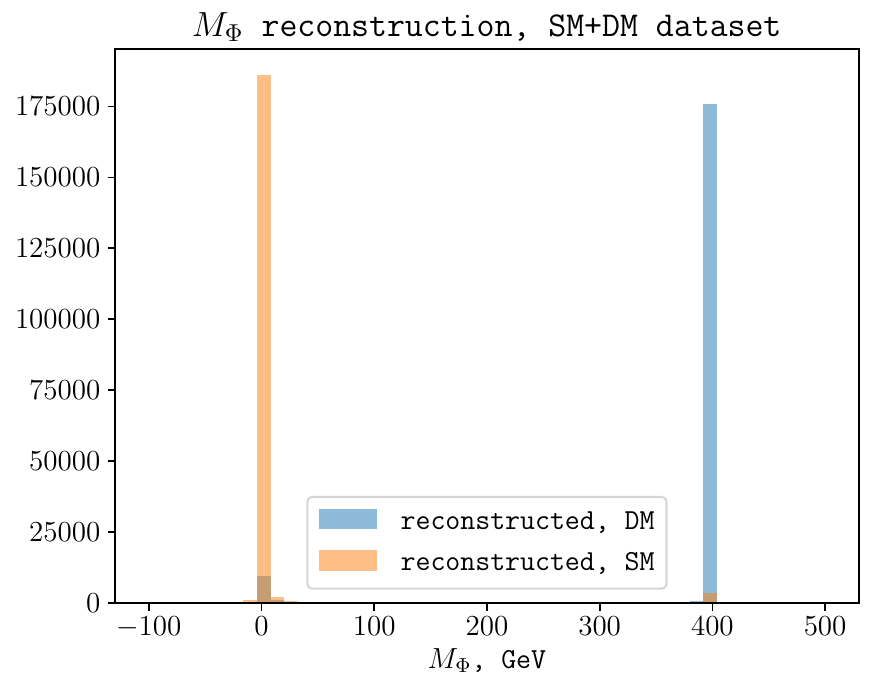}
    \includegraphics[width=.32\textwidth]{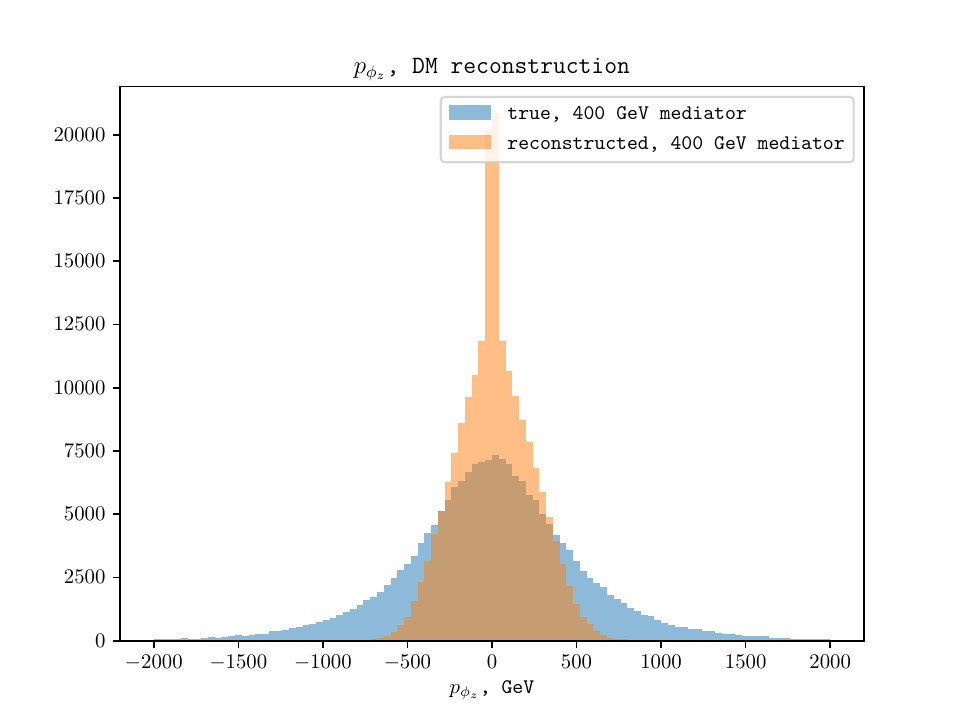} 
\caption{Event distributions for the outputs of the network trained on the 1:1 mix of SM and DM samples. Predicted values shown in blue, target values - in orange. Bottom center figure represents predictions of the same network for 2 samples: SM and DM. 
Orange histogram corresponds to the prediction of the NN on the SM sample, blue - on the DM sample. 
Bottom right figure depicts the reconstructed $p_{\phi_z}$ from the SM+DM NN, where only the DM sample is used for prediction.}
\label{fig:DM_net}
\end{figure}

\begin{figure}[h]
    \centering
    \includegraphics[width=.5\textwidth]{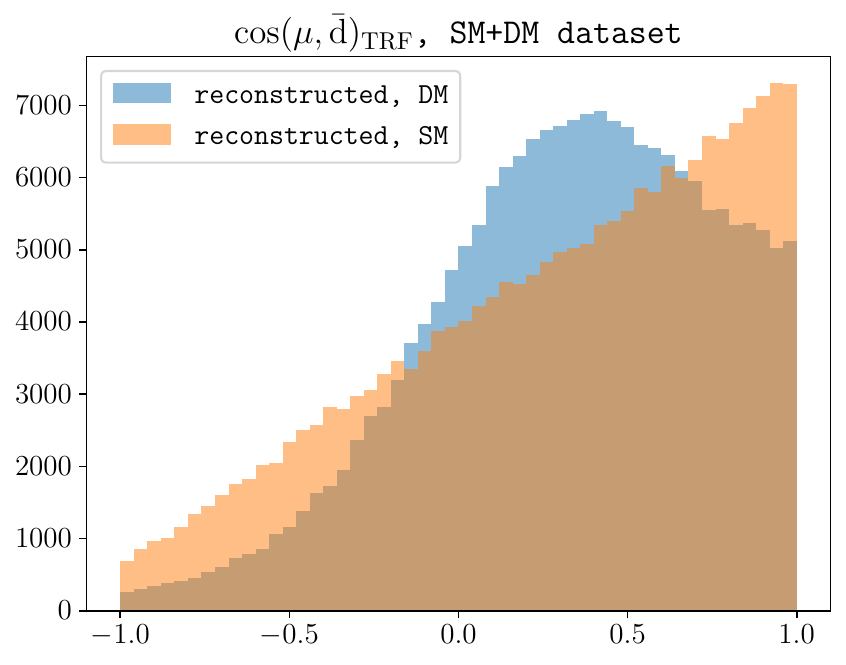} 
\caption{Event distributions for the calculated $cos\theta_{\bar{d} \bar{\ell}}$ using reconstructed 4-momenta from the SM+DM NN. The orange histogram corresponds to the prediction of the NN on the SM sample, blue - on the DM sample.}
\label{fig:DM_net_cos}
\end{figure}

Following the identical approach to the SM NN, the combined SM+DM NN is constructed. The only difference in the architecture of this network compared to the SM NN is the output dimensions. Since this network's main goal is the separation of the DM and SM events, it is trained on 1:1 mix of SM and DM samples. Predictions of the NN on this mix is shown on the Fig.~\ref{fig:DM_net}. 
For the $cos\theta_{\bar{d} \bar{\ell}}$ reconstruction, the same network is used on the SM and DM samples separately. The resulting distributions are depicted in Fig.~\ref{fig:DM_net_cos}. Good agreement is maintained between true and predicted values for $cos\theta_{\bar{d} \bar{\ell}}$ for the SM sample; for the DM, the desired flat distribution is not achieved. Instead, it is more comparable to the distribution shown on the Fig.~\ref{fig:spincorrs_delphes}. This fact could be attributed to the poor reconstruction of the $p_{\phi_z}$ (Fig.~\ref{fig:DM_net}), which, in turn, probably originated from the ambiguity in the sign of $p_{\phi_z}$ from the eq.~\eqref{eq:5}.
Further improvements are possible, such as adapting the architecture described in Refs.~\citen{10.21468/SciPostPhys.14.6.159},~\citen{Sonnenschein:2005ed} for the DM search.

\section{Results}
\label{sec:results}
In this work we have presented a promising new observable to distinguish events containing scalar of pseudoscalar mediator of dark matter based on the angular correlations of the top-quark in SM. The new angular variable discussed in Section~\ref{sec:spincor} provides a significant separation between SM and DM processes. In combination with separation by missing transverse momentum this variable can be used to improve search for dark matter at various colliders. In order to further enhance the separation power of the method two approaches were tested: analytical approach based on the kinematic relations shown in Section~\ref{subsec:analyt} and the machine learning approach using neural networks for the reconstruction of 4-momenta described in Section~\ref{subsec:NN}. Both approaches did not achieve desired improvements; however, the approach using neural networks shows significant potential and, probably, can be improved with more sophisticated NN architectures.
\section*{Acknowledgments}
This study was conducted within the scientific program of the Russian National Center for Physics and Mathematics, section 5 «Particle Physics and Cosmology».
\clearpage
\bibliographystyle{ws-ijmpa}
\bibliography{main}

\end{document}